\documentclass[pdftex]{aa}
\usepackage{graphicx}
\usepackage{natbib}
\usepackage{amssymb,amsmath}
\usepackage{subfigure}
\usepackage[normalem]{ulem}

\usepackage{pslatex}	    		

\bibpunct{(}{)}{;}{a}{,}{,}

\usepackage{ctable,rotating}
\newcommand{\p}{_{\text{p}}}
\newcommand{\eff}{_{\text{eff}}}
\newcommand{\sub}{_{\text{s}}}
\newcommand{\sstar}{_{\star}}
\newcommand{\MM}{\mathcal{M}}
\newcommand{\E}{_{\oplus}}

\begin{document}

\title{Galactic cosmic rays on extrasolar Earth-like planets: I. Cosmic ray flux}

\author{J.--M. Grie{\ss}meier
          \inst{1,2}
          \and
          F. Tabataba-Vakili
          \inst{3}
          \and
          A. Stadelmann
          \inst{4}
          \and
          J. L. Grenfell 
          \inst{5,6}
          \and
          D. Atri
          \inst{7}
 }

\institute{
	LPC2E - Universit\'{e} d'Orl\'{e}ans / CNRS, 3A, Avenue de la Recherche Scientifique, 45071 Orléans cedex 2, France. \email{jean-mathias.griessmeier@cnrs-orleans.fr}
        \and
Station de Radioastronomie de Nan\c{c}ay, Observatoire de Paris - CNRS/INSU, USR 
  704 - Univ. Orl\'{e}ans, OSUC, route de Souesmes, 18330 Nan\c{c}ay, France
        \and
        Atmospheric, Oceanic and Planetary Physics, Department of Physics, University of Oxford, Clarendon Laboratory, Parks Road, Oxford OX1 3PU, UK
        \and
        Technische Universit\"{a}t Braunschweig, Mendelssohnstr. 3, 38106 Braunschweig, Germany
        \and
        Zentrum für Astronomie und Astrophysik (ZAA), Technische Universität Berlin (TUB), Hardenbergstr. 36, 10623 Berlin, Germany
        \and
        now at: Extrasolare Planeten und Atmosph\"{a}ren (EPA), Institut für Planetenforschung, Deutsches Zentrum für Luft- und Raumfahrt (DLR),
	Rutherfordstr. 2, 12489 Berlin, Germany
        \and
	Blue Marble Space Institute of Science, 1200 Westlake Ave N Suite 1006, Seattle, WA 98109, USA
}

\date{Version of \today}

\abstract{
Theoretical arguments indicate that close-in terrestial exoplanets may have weak magnetic fields, especially in the case of planets more massive than Earth (super-Earths). Planetary magnetic fields, however, constitute one of the shielding layers that protect the planet against cosmic-ray particles. In particular, a weak magnetic field results in a high flux of Galactic cosmic rays
that extends to the top of the planetary atmosphere.}
{We wish to quantify the flux of Galactic cosmic rays to an exoplanetary atmosphere as a function of the particle energy and of the planetary magnetic moment.}
{We numerically analyzed the propagation of Galactic cosmic-ray particles through planetary magnetospheres. We evaluated the efficiency of magnetospheric shielding as a function of the particle energy (in the range 16 MeV $\le$ E $\le$ 524 GeV) and as a function of the planetary magnetic field strength (in the range 0 $M_\oplus$ $\le$ M $\le$ 10 $M_\oplus$). Combined with the flux outside the planetary magnetosphere, this gives the cosmic-ray energy spectrum at the top of the planetary atmosphere 
as a function of 
the planetary magnetic moment.
}
{
We find that the particle flux to the planetary atmosphere can be increased by more than three orders of magnitude in the absence of a protecting magnetic field. 
For a weakly magnetized planet ($\MM=0.05\,\MM\E$), only particles with energies below 512 MeV are at least partially
shielded.
For a planet with a magnetic moment similar to that of Earth, this limit increases to to 32 GeV, whereas for a strongly magnetized planet ($\MM=10.0\,\MM\E$), partial shielding extends up to 200 GeV. 
Over the parameter range we studied, strong shielding does not occur for weakly magnetized planets. For a planet with a magnetic moment similar to that of Earth, particles with energies below 512 MeV are strongly shielded, and for strongly magnetized planets, this limit increases to 10 GeV. 
}
{We find that 
magnetic shielding strongly controls the number of cosmic-ray particles reaching the planetary atmosphere.
The implications of this increased particle flux are discussed in a companion article.
}


\titlerunning{Galactic cosmic rays on extrasolar Earth-like planets: I.}
\authorrunning{J.--M. Grie{\ss}meier et al.}

\maketitle

\section{Introduction}

\subsection{Super-Earths}

Since the first discovery of an extrasolar planet around a 
Sun-like star \citep{Mayor95}, the number of known exoplanets has been growing steadily. Advances in instrumentation and analysis have allowed the detection of smaller planets almost from year to year.

The detection of planets with $M\le 10 M\E$, (i.e.,~with a mass smaller than ten terrestrial masses, which is usually considered as the upper limit for super-Earths)
became possible about ten years after the discovery of the first exoplanets around Sun-like stars. 
The first super-Earth detected by radial velocity measurements is GJ 876d, a planet with $\sim\!\!7.5 M\E$ 
orbiting an M-dwarf star \citep{Rivera05}.
Other discoveries of planets with similar masses followed quickly, using a variety of detection techniques. For example, OGLE-2005-BLG-390Lb, a planet with $5.5 M\E$ in an orbit of 5 AU, was found by microlensing 
\citep{Beaulieu06}, and CoRoT-7b, a planet with a radius of 1.7 $R\E$ and a mass of 
$\sim\!\!7-8 M\E$ was detected by transit observations \citep{Leger09,Queloz09,Leger11}.

The detection limit for planets around main-sequence stars has decreased considerably over the last few years: Transit observations have found planets with radii slightly smaller than Earth's \citep[$R=0.87 R\E$,][]{Fressin12}, approximately half of Earth's radius \citep[$R=0.57 R\E$,][]{Muirhead12}\footnote{\citet{Dressing13} label this star as potentially evolved, which would change the derived planetary radius.}, and even a planet with a radius smaller than that of Mercury \citep[$R=0.30 R\E$,][]{Barclay13N}.
At the same time, radial velocity observations have detected planets with masses similar to that of Earth \citep{Dumusque12}.
In total, 123 planets with masses $\lesssim 10 M\E$ are known to date (November 2014), and 5 planets have a mass $\lesssim 1 M\E$ \footnote{Up-to-date numbers can be found at the Extrasolar Planets 
Encyclopaedia: {\tt http://www.exoplanet.eu} \citep{Schneider11}.}. 

\subsection{M star planets}

One of the many fascinating questions in the field of exoplanet studies is the search for 
habitable worlds. 
Because of their relatively small mass, low luminosity, long lifetime, and large abundance in the
Galaxy, M dwarfs are sometimes suggested as prime targets for searches for terrestrial habitable 
planets \citep[see, e.g.,][]{Tarter07, Scalo07}.
For this reason, it is interesting to examine super-Earths orbiting M-dwarf stars.
Currently, 95 of the total of 1850 known exoplanets are located around M-dwarf stars ($M_{\star}\le0.5 M_{\sun}$), and 71 of these are super-Earths.
While this number is currently still limited, 
the detection of cold debris disks around M-dwarf stars seems to indicate that planets may be as
frequent for M stars as they are for F, G, and K stars \citep{Lestrade06}. This expectation was confirmed by recent estimations based on the Kepler Input Catalog \citep{Dressing13}. These estimations indicate that the occurrence rate of planets with a radius $0.5R \E \le R \le 4 R\E$ orbiting an M dwarf in less than 50 days is $0.9^{+0.04}_{-0.03}$ planets per star, and the number of planets per star increases with decreasing planetary mass. 
Thus, the number of currently known super-Earths orbiting M dwarfs can be expected to represent only a tiny fraction of the total number. Qualitatively similar results have been obtained by \citet[][based on KEPLER data]{Howard12} and by \citet[][based on HARPS data]{Bonfils13}.

A necessary, but not sufficient condition for a planet to be considered habitable is the existence of liquid water on its surface \citep[e.g.,][]{Kasting93,Selsis07}.
For M-dwarf stars, the liquid water habitable zone (HZ, the range of distances around a star where conditions allow the presence of liquid water on the planetary surface) is much closer to the star than for Sun-like G-dwarf stars. Its precise location depends on the stellar irradiance (which is a function of the stellar mass), but the outer edge of this region typically is $\le$0.3 AU.  
A first Earth-sized planet has recently been detected in the circumstellar habitable zone of an M-dwarf star \citep{Quintana14}, and 
the number of as yet undetected planets at such a location is expected to be high:
\citet{Dressing13} estimated a mean number of Earth-sized planets in the habitable zone of M dwarfs of $0.15^{+0.13}_{-0.06}$ planets per cool star. Similarly, \citet{Bonfils13} expect  $0.41^{+0.54}_{-0.13}$ planets per M-dwarf star. The difference between the two numbers is, among other factors, due to the different definition of a super-Earth, and the conversion of a minimum radius to a minimum mass. 
The results of \citet{Dressing13} have been reanalyzed with modified HZ limits in \citet{Kopparapu13}, who obtained a frequency of $0.48^{+0.12}_{-0.24}$ and $0.53^{+0.08}_{-0.17}$ terrestrial 
exoplanets per M-dwarf habitable zone for the conservative and optimistic limit of the HZ boundaries, respectively.
In any case, all studies agree qualitatively: Super-Earths in the liquid water habitable zone of M-dwarf stars are very likely to be abundant!

Care should be taken not to confound this ``liquid water habitable zone'' with a zone where all planets will indeed be habitable. A number of additional conditions needs to be fulfilled for ``true'' habitability \citep[see, e.g.,][]{Lammer09AARV, Lammer10AB}.
The close stellar distance and the special characteristics of M dwarfs pose additional problems and constraints to habitability, many of which were extensively reviewed by \citet{Tarter07} and \citet{Scalo07}. 
Intense stellar flares are common, especially for young M-dwarf stars, but the radiation can be
shielded by the planetary atmosphere \citep{Heath99}. Both direct UV radiation
\citep{Buccino07} and indirect UV radiation generated by energetic photons \citep{Smith04} 
can have important consequences for biogenic processes on M-star planets.

A number of additional questions are linked to the presence and intensity of a planetary magnetic field. 
There is increasing evidence indicating that 
super-Earth planets 
are likely to have a weak magnetic field, which has a number of interesting consequences. One of these is the enhanced flux of cosmic-ray particles to the planetary atmosphere, with potential implications ranging from atmospheric chemistry to an increase of the radiation dose close to the planetary surface.

\subsection{Cosmic-ray populations}
\label{sec-populations}

\begin{figure}[tb] \begin{center}
     \includegraphics[width=0.95\linewidth]
{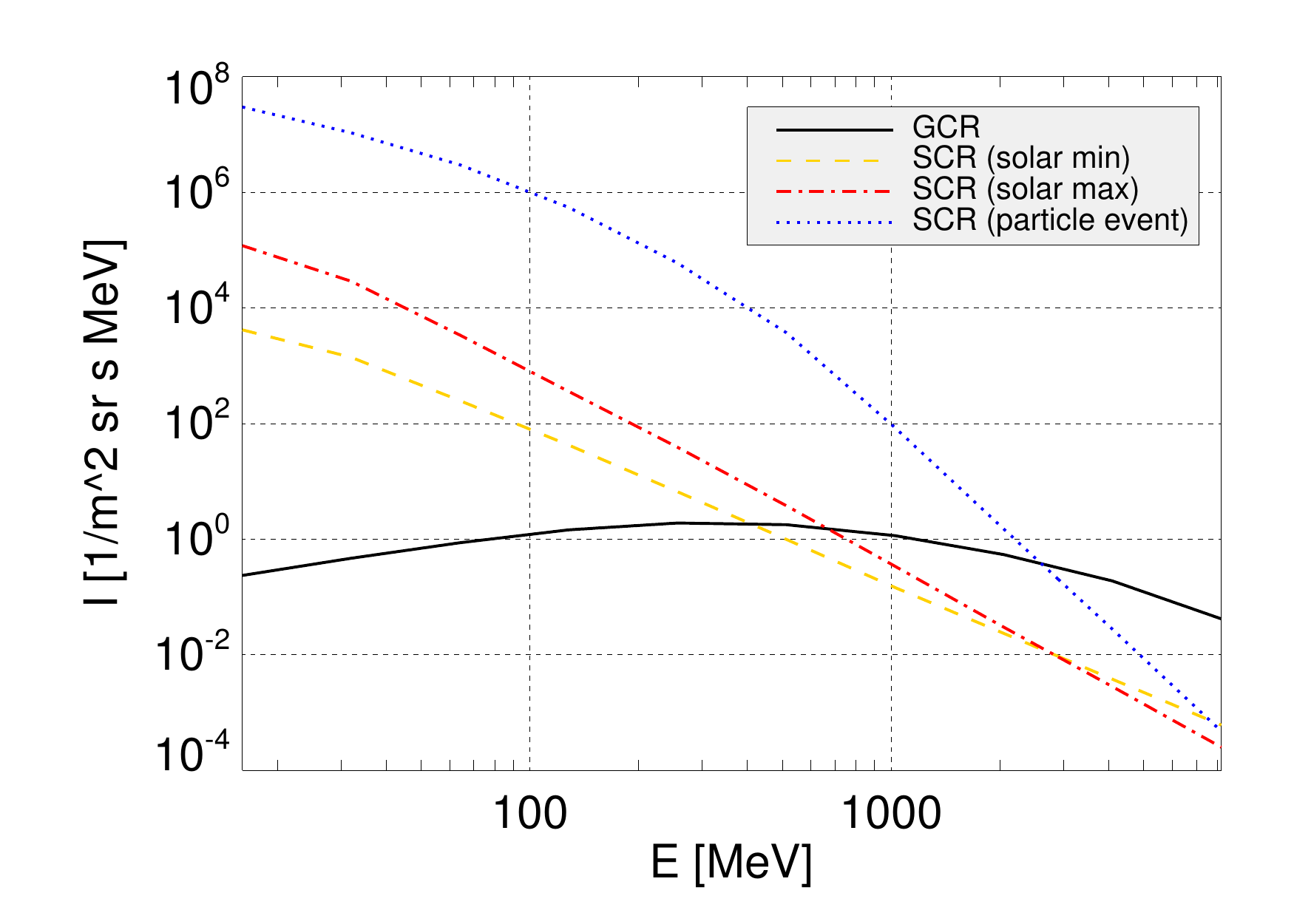}
\caption{Comparison of cosmic-ray spectra for a planet at 0.2 AU. Solid line (black): Galactic cosmic rays. Dashed line (yellow): Average of stellar cosmic rays during activity minimum. Dash-dotted line (red): Average of stellar cosmic rays during activity maximum. Dotted line (blue): Stellar cosmic rays during a stellar particle event.
\label{fig-crspectra}}
\end{center}
\end{figure}

The two main populations of cosmic rays are stellar cosmic rays (generated by the planetary host star) and Galactic cosmic rays (generated by sources outside the planetary system). For a planet, the Galactic cosmic rays flux can be regarded as isotropic and approximately constant (although the flux of low-energy particles is slightly modulated by the solar activity); it also only weakly depends on orbital distance. Outside the Earth's magnetosphere, it has a peak close to 500 MeV, and the flux decreases for both higher and lower particle energies (Fig. \ref{fig-crspectra}, solid line). In contrast, the flux of stellar cosmic rays is time dependent and depends on the stellar activity. 
Figure \ref{fig-crspectra} shows three different cases for stellar cosmic rays (for a planet at 0.2 AU): The average flux during activity minimum (dashed line), the average flux during activity maximum (dash-dotted line), and the cosmic-ray flux during a stellar particle event (dotted line) \citep[see][for details concerning the stellar cosmic ray spectra]{Grenfell12,TabatabaVakili15}.
In either case, the flux of stellar cosmic rays is high for low particle energies and  steeply decreases for higher energies. 
At low particle energies, solar cosmic rays dominate over Galactic cosmic rays. For Earth, the energy where both contributions have similar fluxes is between $\sim$100-1000 MeV (depending on solar activity). For higher energies, Galactic cosmic rays are dominant. To compare the relative strength of the two cosmic-ray contributions at other planets, one has to take into account the planetary orbital distance (which has a strong influence on stellar cosmic rays, but only weakly affects Galactic cosmic rays). For example, for a planet at 0.2 AU, the energies of comparable fluxes are 400 MeV, 660 MeV, and 2.6 GeV for our three cases of stellar activity (see Fig. \ref{fig-crspectra}). The different energy ranges of stellar and Galactic cosmic rays also mean that the particles have different penetration depths in the atmosphere, where they encounter different atmospheric species in different concentrations. For this reason, stellar and Galactic cosmics rays are usually treated separately.  

The effect of stellar cosmic rays on exoplanetary atmospheres has been addressed by \citet{Grenfell12}. 
An updated analysis of this case is presented in a companion article \citep{TabatabaVakili15}.

In the current work, we focus on the effect of Galactic cosmic rays on exoplanets. We will, however, return to the comparison between stellar and Galactic cosmic rays in Sect. \ref{sec-I}.

\subsection{Cosmic rays on super-Earths orbiting M star}

The severity of the effects generated by cosmic rays depends on a number of physical parameters. Previous work has studied the influence of
stellar age
\citep[which is a defining variable for the stellar wind parameters, see][]{Griessmeier05AB},
the influence of
 the orbital distance \citep[][]{Griessmeier09}, 
the influence of
 the presence or absence of tidal locking through its influence on the planetary
magnetic field \citep[][]{Griessmeier05AB,Griessmeier09} and 
the influence of
 the planetary type
\citep[again, through the estimated magnetic field,][]{Griessmeier09}. 
Using these results, the effects of Galactic cosmic rays on the planetary atmospheric chemistry were calculated \citep{Grenfell07AB}, and surface radiation doses were evaluated \citep{Atri13}.
Here, we take a 
more general approach
and systematically study the influence of the planetary magnetic field: Instead of
applying a model for the planetary magnetic moment, we show how magnetic protection
varies as a function of the planetary magnetic dipole moment.

Including this shifted focus, the main differences with respect to previous work \citep{Griessmeier05AB,Griessmeier09} 
are the following:
\begin{itemize}
        \item   Here, we systematically study the influence of the planetary
                magnetic field. Instead of applying a model for the planetary
                magnetic moment, we show how magnetic protection varies as a
                function of the planetary magnetic dipole moment.
        \item   We have included low-energy cosmic-ray particles down to
                16 MeV.
        \item   We have included high-energy cosmic-ray particles up to
                524 GeV. Thus, the total energy span of the study is ($16$ MeV
                $\le E \le 524$ GeV). The limits of the previous studies were 64 MeV to 8.192 GeV.
                We have thus decreased the minimum energy by a factor of 4 and 
                increased the maximum energy by a factor of 64.
        \item   The calculation of high-energy particles made it necessary to
                multiply the number of 
                particles by a factor of 4 (to 28 million particles per configuration and energy). 
        \item   We have extended the analysis of the incoming particle 
                population: We 
                discuss the magnetospheric filter function and compare particle 
                energies and rigidities. We also calculate the (magnetic-moment dependent) 
                energy above which a significant fraction of the
                particles reach the atmosphere, and evaluate the energy of maximum particle flux.
\end{itemize}   

\begin{figure}[tb] \begin{center}
     \includegraphics[width=0.95\linewidth]
{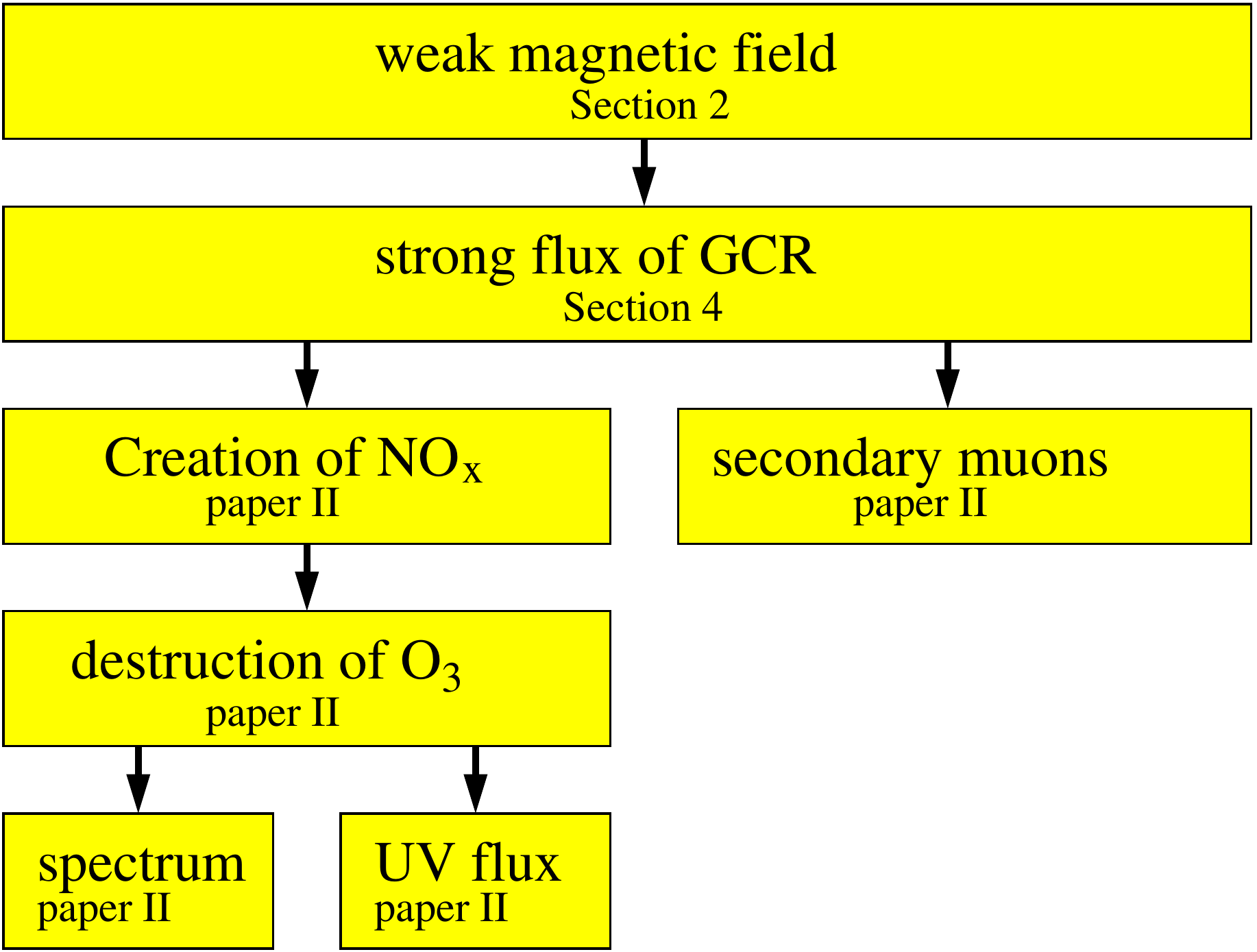}
\caption{Effects of Galactic cosmic rays discussed in this work and in the companion article (paper II).
\label{fig-plan}}
\end{center}
\end{figure}

This paper is organized as follows (see also Fig. \ref{fig-plan}):
Section \ref{sec-planet} describes the types of planetary situations we are interested in, namely close-in terrestrial planets around M-dwarf stars with weak magnetic fields (Sect. \ref{sec-magmoment}).
In Sect. \ref{sec-scenarios}, we present the planetary parameters used in our calculations.
Section \ref{sec-model} decribes the models and numerical tools we use: 
The stellar wind model is presented in Sect. \ref{sec-stellarwind}, and the planetary magnetospheric model is described in Sect. \ref{sec-magnetosphere}.
The cosmic-ray model is explained in 
Sect. \ref{sec-cr}. 
The Galactic cosmic-ray fluxes are analyzed in Sect. \ref{sec-results}.
We have also re-evaluated the implications of such high Galactic cosmic ray fluxes on planetary atmospheres and surfaces. These 
implications are discussed in a companion article \citep[][hereafter ``paper II'']{Griessmeier15inprep2}.
Section \ref{sec-conclusions} closes with some concluding remarks.

\section{Planetary situation} \label{sec-planet}

\subsection{Planetary magnetic moment} \label{sec-magmoment}

Until recently, the magnetic field of extrasolar rocky planets was not only unaccessible to observations, but also to theoretical studies. This has changed in recent years, with a number of studies investigating whether and under which conditions a super-Earth can host a significant magnetic field. While these studies use very different approaches, they all come to a similar conclusion: Magnetic fields on super-Earths around M-dwarf stars are likely to be weak and  short-lived in the best case, or even nonexistent in the worst case. The relevance of such fields and their potential detectability is discussed elsewhere \citep[][]{Griessmeier14inbook}.
With this in mind, the question of planetary magnetic shielding against Galactic cosmic rays becomes indeed important.

In previous studies \citep{Griessmeier05AB,Griessmeier09}, we have attempted to estimate the magnetic moment of an Earth-like exoplanet around a K/M-dwarf star. Based on analytical scaling laws, we found the planetary magnetic dipole moment to lie in the range 
$0.02 \mathcal{M}\E \le \mathcal{M} \le 0.15 \mathcal{M}\E$, where $\mathcal{M}\E$ is 
the value of Earth's current magnetic moment. We then adopted the maximum value of $0.15 \mathcal{M}\E$ to obtain a lower limit for the cosmic-ray flux to the atmosphere.

However, such simplified quantitative estimates of magnetic fields are not very reliable. More complex approaches, however, yield values which are not only model dependent, but also depend on the precise planetary parameters. For this reason, we take a different approach here:
Instead of applying a model for the planetary magnetic moment, we show how magnetic protection
varies as a function of the planetary magnetic dipole moment. 
We thus defer the evaluation of the planetary magnetic field, and calculate a large number of representative cases instead. This allows us to systematically study the influence of the planetary magnetic field on the flux of Galactic cosmic rays to the planet. 
In this way, we explore the range $0.0 \,\mathcal{M}\E \le \mathcal{M} \le 10.0 \,\mathcal{M}\E$ for the magnetic moment, and the range of $16 \,\text{MeV} \le E \le \,524 \, \text{GeV}$ for the particle energy.

\subsection{Planetary scenarios}
\label{sec-scenarios}

As mentioned above, previous work has focused on the dependence of planetary magnetic shielding on the stellar age
\citep[which is a defining parameter for the stellar wind parameters, see][]{Griessmeier05AB}, on the orbital distance \citep[][]{Griessmeier09}, on the presence of absence of tidal locking through its influence on the planetary
magnetic field \citep[][]{Griessmeier05AB,Griessmeier09} and on the planetary type
\citep[again, through the estimated magnetic field][]{Griessmeier09}. 
Here, we take a slightly different approach and treat the planetary magnetic field $\mathcal{M}$ as a free parameter, keeping the other parameters fixed.

As shown in Table \ref{tab-parameters}, the following parameters are kept fixed:
The stellar mass $M\sstar$ ($M\sstar=0.5 M_{\sun}$), the stellar radius $R\sstar$ ($R\sstar=0.46 R_{\sun}$), the stellar age $t\sstar$ (4.6 Gyr), and the orbital distance $d$ ($d=0.2$ AU). These parameters determine the stellar wind (which thus remains unchanged, too). 
In addition, we keep constant the planetary parameters 
of mass $M\p$ ($M\p=1.0 M_{\oplus}$) and radius $R\p$ ($R\p=1.0 R_{\oplus}$).
The only planetary or stellar parameter we vary in the present study is the planetary magnetic dipole moment $\mathcal{M}$ (see Table \ref{tab-parameters}). 

We note that the precise values of $M\sstar$, $R\sstar$ and of $d$ do not matter much for the amount of cosmic rays reaching the planetary atmosphere.
In particular, using $M\sstar=0.45 M_{\sun}$, $R\sstar=0.41 R_{\sun}$ and $d=0.153$ AU (the values used in the atmospheric model of paper II) rather than the values given in Table \ref{tab-parameters} leads to identical particle fluxes within the numerical error.
The precise value of $t\sstar$, however, can have some limited influence in some cases.
Based on the results of \citet{Griessmeier09} and \citet{Griessmeier05AB}, the influence of the parameters $d$, $M\sstar$, $R\sstar$, and $t\sstar$ is discussed in more detail
in Sect. \ref{sec-Ecrit}.
The main goal of this study, however, is to analyze
how the flux of cosmic rays at the top of the atmosphere
varies as a function of the planetary magnetic dipole moment $\mathcal{M}$ and particle energy $E$ over the ranges given in Table \ref{tab-parameters}.

For the planetary magnetic moment $\mathcal{M}$, we allow values between 0 and 10 times the terrestrial value (see Table \ref{tab-parameters}). The minimum value of the magnetic dipole moment in this study is 0. This corresponds to an   unmagnetized planet, for which we take the cosmic-ray energy spectrum outside the Earth's magnetosphere [\citet[][]{Seo94} for $E\le$ 8 GeV, and \citet[][using their ``median'' case]{Mori97} for $E\ge$ 16 GeV].
The maximum value of the magnetic dipole moment in this study is ten times the present Earth value, which corresponds to an extremely strongly magnetized planet.

Table \ref{tab-parameters} also shows the energy range we use for cosmic-ray protons. The range of particle energies was chosen such that all particles relevant for the processes described in \citet{Griessmeier15inprep2} are included (the lowest and the highest energy particles do not contribute significantly). 
To serve as input for future studies (for which the relevant energy range cannot be known), we extended the energy range beyond the values required for \citet{Griessmeier15inprep2}.

\begin{table}[!h]
\begin{center} 
   \caption[Parameters for cosmic ray calculation]
   {Parameters for the different planetary configurations.}
   \begin{tabular}{l c}\hline \hline
        parameter
                & value
                \\[4pt] \hline 
        $M\sstar$ [$M_{\sun}$]              
                & $0.5^\dag$
                \\[4pt] 
        $R\sstar$ [$R_{\sun}$]               
                & $0.46^\dag$
                \\[4pt] 
        $t\sstar$            
                & 4.6 Gyr$^\ddag$
                \\[4pt] 
        $d$ [AU]            
                & $0.2^\dag$
                \\[4pt] 
        $M\p$ [$M_{\oplus}$]              
                & 1.0
                \\[4pt] 
        $R\p$ [$R_{\oplus}$]               
                & 1.0
                \\[4pt] 
        $\mathcal{M}$            
                & $0.0 \,\mathcal{M}\E \le\mathcal{M} \le10.0 \,\mathcal{M}\E$        
                \\[4pt] 
        $E$             
                & $16 \,\text{MeV} \le E \le \,524 \text{GeV}$       
                \\[4pt] \hline
   \end{tabular}
   \tablefoot{$M\sstar$: stellar mass, $R\sstar$: stellar radius, 
   $t\sstar$: stellar age, 
   $d$: planetary orbital distance, 
   $M\p$: planetary mass, $R\p$: planetary radius, 
   $\mathcal{M}$: planetary magnetic moment, $E$: particle energy. \\
   Notes: $^\dag$: These values have previously been shown to have little influence on cosmic-ray shielding (see text). 
   $^\ddag$: This value has previously been shown to have some influence on cosmic-ray shielding (see text).
   }
\label{tab-parameters}
\end{center} 
\end{table}

\section{Models}
\label{sec-model}

In this part, we describe the models used throughout this article. 
In Sect. \ref{sec-stellarwind}, we describe the stellar wind model that we used to model the 
planetary magnetosphere (Sect. \ref{sec-magnetosphere}).  
In Sect. \ref{sec-cr}, 
we describe the model used to calculate the flux of Galactic cosmic rays through this planetary magnetosphere.
The results obtained with these models are given in Sect. \ref{sec-results}.

\subsection{Stellar wind model}\label{sec-stellarwind}

It is expected that at the close orbital distances of M-star habitable zones the stellar wind has 
not yet reached the quasi-asymptotic velocity regime.  
Because of the low stellar wind velocity, the planetary magnetosphere is less
strongly compressed at such distances than one would expect from stellar wind
models with a constant velocity.
To capture this behavior realistically 
and to correctly describe the flux of Galactic cosmic-rays into the atmospheres of close-in exoplanets, we
require a model with a radially dependent stellar wind velocity. In this work, we chose a stellar wind model based on the 
work by \citet{Parker58}, which adequately describes the 
stellar wind around slowly rotating stars. As our study is limited to stars of solar age, this approximation is justified. 
The model takes as input the stellar mass, radius, and age. It yields a self-consistent solution for the stellar wind velocity as a function of orbital distance $v(d)$, the stellar wind density as a function of the orbital distance $n(d)$, and the stellar wind temperature $T$, which we used as input parameters in the following section.
The details of the stellar wind model are described in \citet{Griessmeier05AB} and \citet{Griessmeier09}.

\subsection{Magnetospheric model}\label{sec-magnetosphere}

The magnetosphere was modeled as a cylinder topped by a half-sphere 
\citep{Voigt81,Voigt95,Griessmeier04,Griessmeier05AB,Stadelmann10}. 
A closed magnetosphere was assumed, that is,~field lines are not
 allowed to pass
through the magnetopause.
With this model, the magnetic field is defined for any point inside 
the magnetosphere as soon as the planetary magnetic moment and the size of the
magnetosphere are prescribed. 
As described in Sect. \ref{sec-planet}, the magnetic moment 
was varied in the range $0.0 \,\mathcal{M}\E \le\mathcal{M} \le10.0 \,\mathcal{M}\E$.
The only ingredient missing to define the magnetospheric magnetic field thus is the size of the magnetosphere.

The size of the
magnetosphere is characterized by the magnetopause standoff distance $R\sub$, that is, the
extent of the magnetosphere along the line connecting the star and the planet.
$R\sub$ can be obtained from the pressure equilibrium at the
substellar point. 

This pressure balance includes the stellar wind ram pressure, the stellar wind
thermal pressure of electrons and protons, and the planetary magnetic field pressure:
\begin{equation}
        m n v\eff^2+2 \, nk_BT=
        \frac{\mu_0f_0^2\mathcal{M}^2}{8\pi^2 R\sub^6}. \label{eq:pressureequilibrium}
\end{equation}
Here, 
$v\eff$ is the effective stellar wind velocity relative to the planet
                ($v\eff=\sqrt{v^2+v^2_\text{orbit}}$), $v$, $n,$ and $T$ are the stellar wind velocity, density, and temperature (which can be taken from Sect. \ref{sec-stellarwind}), 
and $f_0=1.16$ is the form factor of the magnetosphere and includes the magnetic field caused
by the currents flowing on the magnetopause \citep{Voigt95,Griessmeier04}. 
$\mathcal{M}$ is the planetary magnetic dipole moment.

We note that Eq.~(\ref{eq:pressureequilibrium}) does not include a contribution from the stellar wind magnetic pressure on the left-hand side.
In our case (e.g., $d=0.2$ AU), $\vec v$ and $\vec B_{\text{imf}}$ are approximately 
parallel to the line connecting the star and the planet.
\citet{Petrinec97} analyzed the position of the magnetopause for different orientations of the
upstream interplanetary magnetic field. For magnetic fields 
$\vec B_{\text{imf}}$ parallel to the stellar wind flow $\vec v$, 
they found that the substellar standoff distance is not influenced
by the magnetic field. 
Hence, in our case, the contribution of the interplanetary field to the stellar wind pressure is negligible.%
\footnote{There are indeed cases where the stellar magnetic pressure 
may be more important than
the stellar wind ram pressure: (1) For very small orbital distances, the planet may develop an ``ahead shock'' in the notation of \citet{Vidotto10ApJ}; here, however, we are in the case of a ``dayside shock'', where the stellar wind ram pressure dominates. (2) One can use the magnetic pressure to try to evaluate the size of the magnetopause at its flanks rather than at the nose \citep{Vidotto13AA}. 
For our case \citep[orbital distance of 0.2 AU and using the stellar magnetic field model of][]{Griessmeier07AA}, this approach 
yields magnetopause distances much larger than the standoff distance (i.e., the pressure balance at the nose, using the solar wind ram pressure).
To model the magnetosphere, we thus have to use the substellar standoff distance, i.e., its size at the nose rather than at the flanks, and have to use the ram pressure.
This shows that for each situation, one has to carefully check which terms are relevant in the pressure balance.}

From the pressure equilibrium Eq.~(\ref{eq:pressureequilibrium}) the standoff distance 
$R\sub$ is given by\begin{equation}
         R\sub = 
         \left[ \frac{\mu_0f_0^2\mathcal{M}^2}
         {8\pi^2 \left(m n v\eff^2+2 \, nk_BT\right)} \right]^{1/6}.
         \label{eq:Rs}
\end{equation}
The magnetopause standoff distance thus
is a function of the stellar wind conditions (obtained in Sect. \ref{sec-stellarwind}) and of the planetary magnetic moment $\mathcal{M}$. Thus, for each value of the magnetic moment, a different magnetospheric configuration is calculated
according to the cylinder-plus-half-sphere model of \citet{Stadelmann10}. With this, the magnetic field is determined for all points in the magnetosphere. This magnetic field serves as input in the following section.

\subsection{Cosmic-ray calculation}
\label{sec-cr}

To quantify the protection of extrasolar Earth-like planets against Galactic cosmic
rays, the motion of Galactic cosmic protons through the planetary magnetospheres 
was investigated numerically.  
Because no solution in closed form exists, this is only possible through the numerical
integration of many individual trajectories \citep{Smart00}.
For each particle energy in the energy range
16 MeV $\le E \le$ 524 GeV 
and for each
magnetospheric configuration, over 28 million trajectories were calculated, 
corresponding to protons with different starting positions and starting velocity directions.
The particles are launched from the surface of a sphere centered on the 
planet with a radius $r\ge R\sub$, that is,~the particles 
are launched outside the magnetosphere (except for those arriving from the tailward direction).
As compared to our previous studies \citep{Griessmeier05AB,Griessmeier09}, this corresponds to an increase of the number of numerical particles 
per configuration by a factor of four. This increase has become necessary because the high-energy particles 
required a better statistical basis. 
As in any cosmic-ray tracing algorithm, the computing-intensive part is not the
calculation of the particle trajectories, but the evaluation of a complicated
magnetic field for each particle position. For a specific case, \citet{Smart00} estimated that the
magnetic field calculation takes 90\% of the total CPU time.

Once the particles enter the magnetosphere, their motion is 
influenced by the local value of the planetary magnetic field.
This magnetic field is calculated from the 
magnetospheric model determined in Sect. \ref{sec-magnetosphere}, using the particle's instantaneous position.
The trajectories
were calculated using the numerical leapfrog method \citep{Birdsdall85}.
For each energy and magnetic configuration, we counted all particles that reach the atmosphere, described by a
spherical shell one hundred kilometres above the planetary surface
(i.e.,~$R_\text{a}=R\p+100$ km).
More details on the numerical calculation of the cosmic-ray 
trajectories can be found in \citet{Stadelmann10}.

The cosmic-ray flux reaching the top of the atmosphere 
is described in Sect. \ref{sec-results}.
In particular, the flux of particles to the planetary atmosphere is quantified by the filter function $\eta(E,\mathcal{M})$
and by the energy spectrum $I(E,\mathcal{M})$, which are defined in Sects. \ref{sec-eta} and \ref{sec-I},
respectively.

\section{Results: Cosmic-ray flux}
\label{sec-results}

We used the numerical model described above (Sect. \ref{sec-cr}) to calculate the flux of Galactic cosmic rays to the planetary atmosphere.
Previously, the cosmic-ray flux to the atmosphere of a weakly magnetized ($\mathcal{M}=0.15\, \mathcal{M}_{\oplus}$) Earth-like exoplanet orbiting 
a K/M-type star with $M_{\star}=0.5 M_{\sun}$ at a distance of $d=0.2$ AU has been calculated \citep{Griessmeier05AB,Griessmeier09}. It has been shown that such a 
planet will be
subject to very high cosmic-ray fluxes when compared to Earth.
For particle energies below 200 MeV, the 
cosmic-ray flux to the exoplanet was found to be up to one order of magnitude higher than on
Earth, and for energies above 2 GeV, 
magnetospheric shielding is negligible for this exoplanet case. 
As shown by \citet{Griessmeier09}, the effect responsible for this reduced shielding efficiency 
is not the compression of the planetary magnetosphere by the enhanced stellar wind ram pressure
at small orbital distances. 
Instead, the enhanced particle flux is a consequence of the assumed weak planetary magnetic dipole moment ($\MM=0.15\MM_\oplus$, which was estimated assuming tidal locking). Here, we study the question how the shielding efficiency
varies as a function of the planetary magnetic moment.

For this, we first introduce the magnetospheric filter function $\eta(E,\MM)$ in Sect. \ref{sec-eta}.
We then analyze different regimes of magnetospheric shielding in Sect. \ref{sec-Ecrit}. Finally, in Sect. \ref{sec-I}, we evaluate the flux of Galactic cosmic rays to the top of the planetary atmosphere.

\subsection{Magnetospheric filter function $\eta(E,\MM)$}
\label{sec-eta}

In the magnetospheric model, the magnetospheric filter function against cosmic rays is determined in the following way: For each value of the 
particle energy $E$ and planetary magnetic dipole moment $\mathcal{M}$, the number $n_\text{shielded}(E,\MM)$ of 
particles reaching the planetary atmosphere is registered. 
This value is compared to the number $n_\text{unshielded}(E,\MM)$ of
particles reaching the atmosphere of an 
identical, but unmagnetized planet. The magnetospheric filter function $\eta(E,\MM)$ is then defined as
\begin{equation}
        \eta(E,\MM)=
        \frac{n_\text{shielded}(E,\MM)}{n_\text{unshielded}(E,\MM)}. \label{eq-eta}
\end{equation}

\begin{figure}[tb] \begin{center}
     \includegraphics[width=0.95\linewidth]
{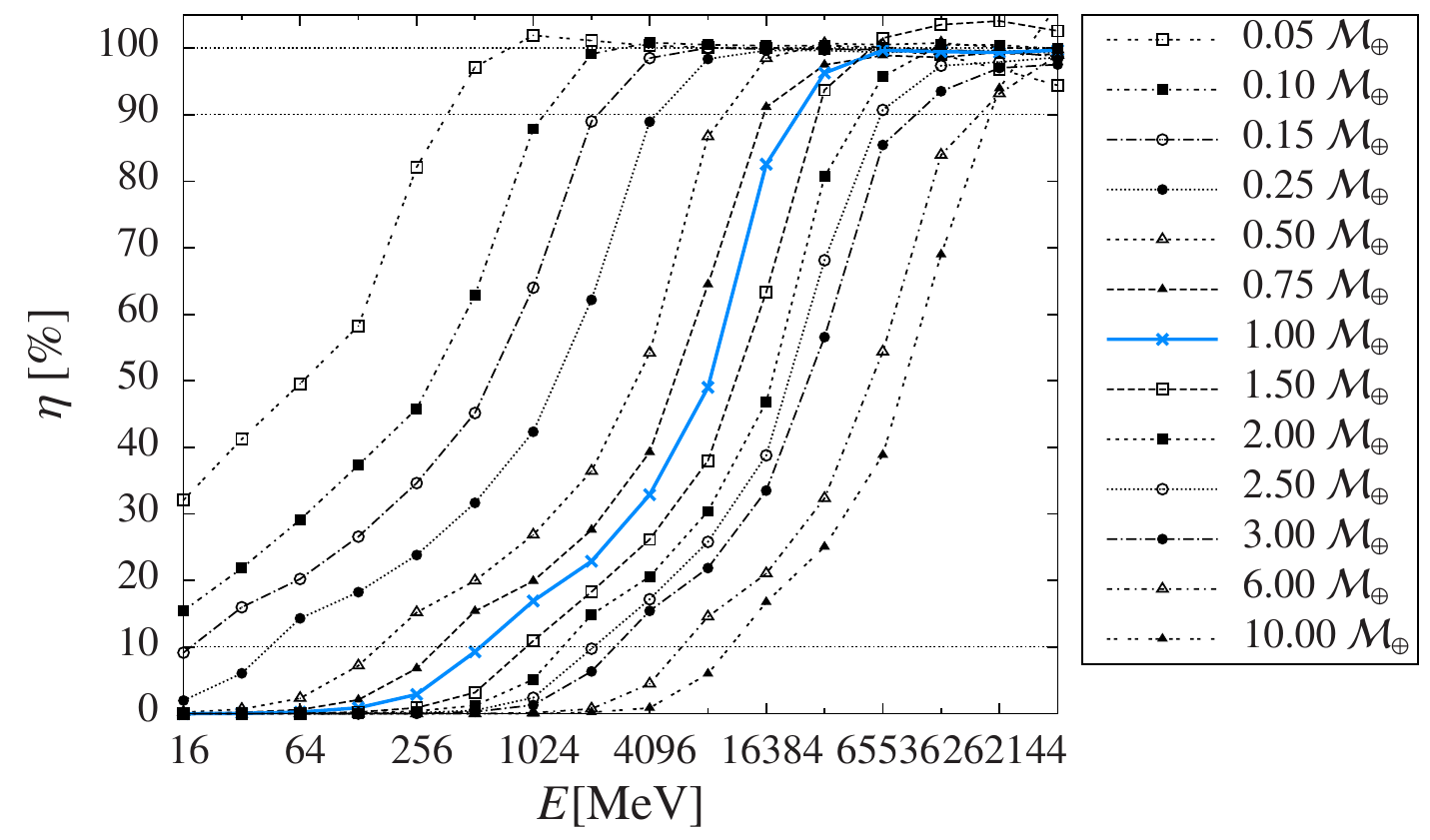}
\caption{Magnetospheric filter function $\eta(E,\MM)$. Blue line: case of a planet with a magnetic moment identical to that of Earth ($\MM=1.0\,\MM\E$).
\label{fig-eta}}
\end{center}
\end{figure}

Figure \ref{fig-eta} shows the dependence of $\eta$ on $E$ and $\MM$ based on the results of our numerical cosmic-ray model (Sect. \ref{sec-cr}). The case of a planet with a magnetic moment identical to that of Earth ($\MM=1.0\,\MM\E$) is shown in blue. 
We define three regimes: ``well shielded'' for $\eta\le0.10$, ``partially shielded'' for $0.10 <\eta<0.90$, and ``unshielded'' for $\eta\ge 0.90$.
One can make the following observations: 
\begin{itemize}
        \item   For a weakly magnetized planet ($\MM=0.05\,\MM\E$), the shielding is very weak. 
                At all energies examined in this study ($E \ge 16$ MeV), some protons (at least 30\%) may 
                penetrate the magnetosphere. Up to 512 MeV, partial shielding prevails.  
                Above 512 MeV, virtually all particles may enter.
        \item   For a planet with $\MM=0.25\,\MM\E$, 
                cosmic rays below 64 MeV are shielded from the atmosphere. 
                Above 4 GeV, all particles may enter the atmosphere.
        \item   For a planet with a magnetic moment similar to that of Earth, shielding is almost 
                perfect for $E\le 512$ MeV. For particle energies $512$ MeV $< E <$ 32 GeV, the   
                planet is partially shielded. Shielding is virtually absent for energies 
                $E\ge 32$ GeV. 
                As described in Sect. \ref{sec-Ecrit}, this case is also
                representative for the situation on present-day Earth.
        \item   For a strongly magnetized planet ($\MM=10.0\,\MM\E$), magnetic shielding is 
                strong for $E\le 10$ GeV. For $10$ GeV $< E < 200$ GeV, cosmic-ray protons are 
                partially shielded. Only for extremly high energies ($E\ge200$ GeV) is the planet 
                unshielded. 
\end{itemize}

\subsection{Magnetospheric shielding regimes}
\label{sec-Ecrit}

\begin{figure*}
\begin{center}
        \subfigure[Magnetospheric filter function as a function of particle 
        energy and planetary magnetic moment.]
        {\includegraphics[angle=0,width=0.45\linewidth]
{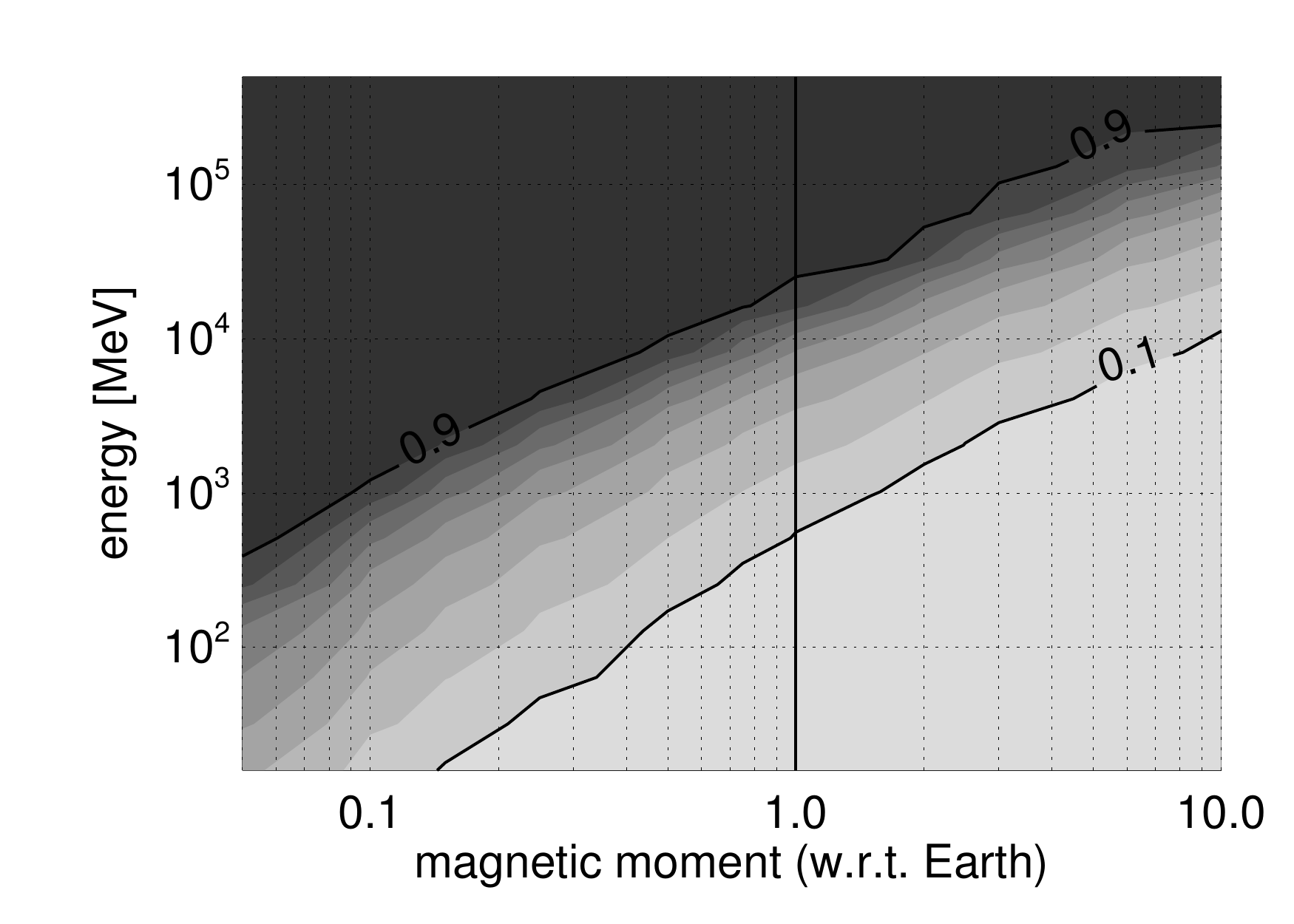}
        \label{fig-limits-energy}}
        \subfigure[Magnetospheric filter function as a function of particle 
        rididity and planetary magnetic moment.]
        {\includegraphics[angle=0,width=0.45\linewidth]
{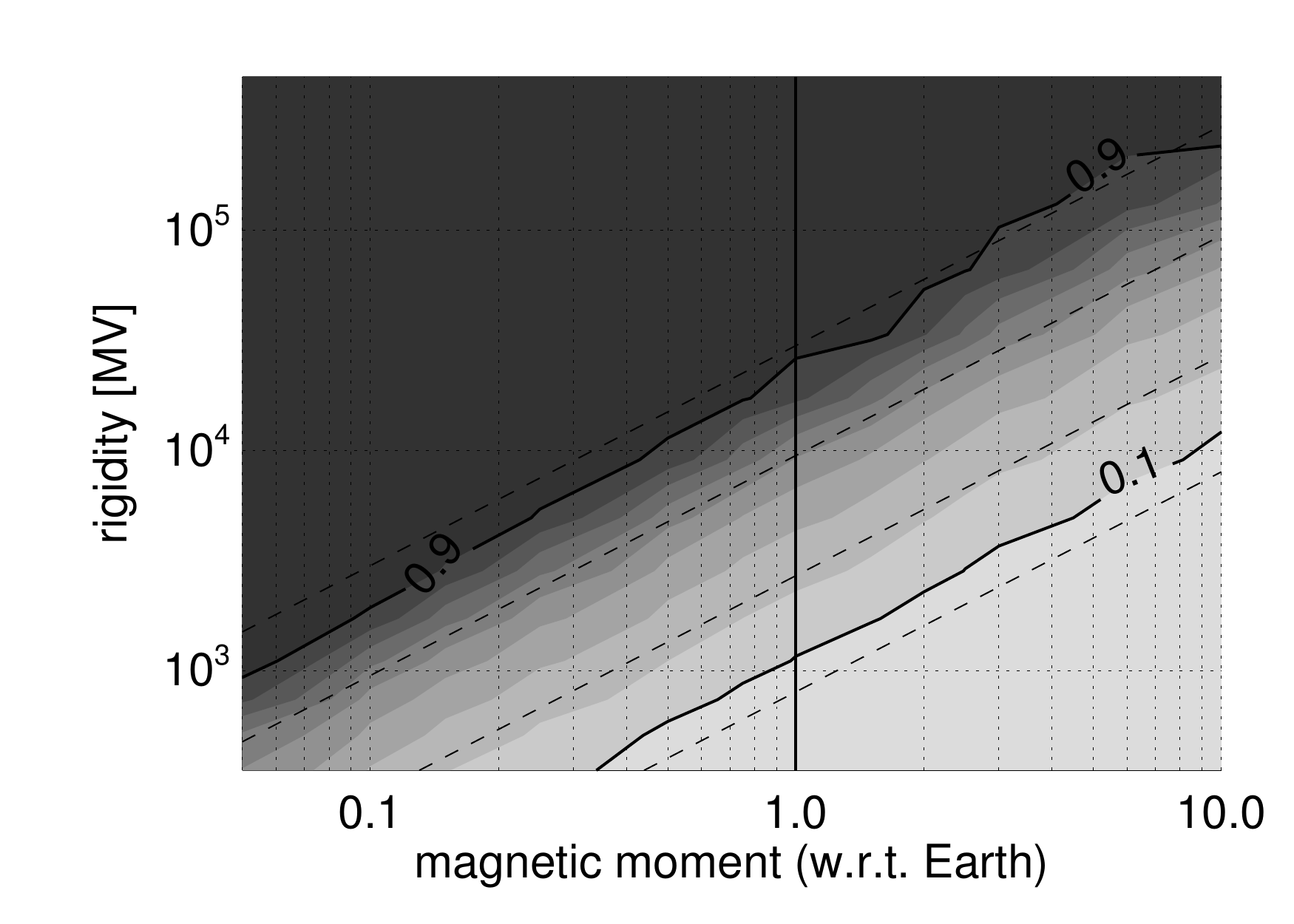}
        \label{fig-limits-rigidity}}
\caption{Magnetospheric filter functions $\eta(E,\MM)$ and $\eta(R,\MM)$ as a function of particle energy $E$ (left panel) or rigidity $R$ (right panel) and planetary magnetic moment $\MM$. Thick lines: $\eta=0.9$ and $\eta=0.1$.
Region above $\eta=0.9$: unshielded. Region between $\eta=0.1$ and $\eta=0.9$: partially shielded. Region below $\eta=0.1$: fully shielded.
Dashed lines: $R\propto \MM$ (see text). 
\label{fig-limits}}
\end{center}
\end{figure*}

To further analyze the result of 
Fig. \ref{fig-eta}, we display $\eta(E,\MM)$ in a contour plot, Fig. \ref{fig-limits-energy}. For each planetary magnetic moment $\mathcal{M}$, we define the critical energies $E_{\text{10\%}}(\MM)$ and $E_{\text{90\%}}(\MM)$ as the energies at which $\eta=0.10$ and $\eta=0.90$, respectively, shown as solid lines in Fig. \ref{fig-limits-energy}.
If we look at the three regimes defined in the previous section, we find the planet ``well shielded'' against particles with $E\le E_{\text{10\%}}$, ``partially shielded'' for 
particles with $E_{\text{10\%}} <E<E_{\text{90\%}}$, and ``unshielded'' for particles with $E \ge E_{\text{90\%}}$. 
Thus, the curves defined by $E_{\text{10\%}}(\MM)$ and $E_{\text{90\%}}(\MM)$ separate the three regimes.
The observations of Sect. \ref{sec-eta} remain valid, of course, but contours allow a more visual display of the three regimes for $\eta(E,\MM)$ and of the critical energies $E_{\text{10\%}}(\MM)$ and $E_{\text{90\%}}(\MM)$.

To compare the behavior of different cosmic-ray particles, it is sometimes useful to use their rigidity $R$ instead of their kinetic energy $E$.
The kinetic energy $E$ (expressed in MeV) of a particle with rest mass $m_0$
can be converted into rigidity $R$ (expressed in MV) using \citet[][Eqs.~(11)-(13)]{Vogt07}:
\begin{equation}
        \frac{R}{\text{MV}}=\sqrt{
        \left(\frac{m_0 c^2}{\text{MeV}}+\frac{E}{\text{MeV}}\right)^2
        -\left(\frac{m_0 c^2}{\text{MeV}}\right)^2
        }.
        \label{eq-R}
\end{equation}
In this, the rigidity $R= p/q$ is 
the (relativistic) momentum of the particle divided by its charge. This choice of variable is useful for one specific reason: 
If one looks at the equation of motion of a particle in a magnetic field, 
the only parameter of the particle that enters this equation is the rigidity $R$. Thus, particles of the same rigidity behave in the same way.
 
Moreover, for a cosmic-ray particle in a planetary magnetopause, a change in particle rigidity $R$ can, in some cases, have exactly the same effect as a change in magnetospheric configuration, characterized by the magnetic moment $\MM$ and the stellar wind ram pressure $p_\text{sw}$ \citep[][]{Vogt07}. This is 
the case especially for two locations, namely (a) close to the magnetopause, and (b) the inner magnetosphere close to the planet. For these cases, we compare two configurations, denoted by subscripts 1 and 2. The first configuration is thus characterized by $\MM_1$ and $p_\text{sw,1}$, and the second configuration by $\MM_2$ and $p_\text{sw,2}$.
\begin{itemize}
\item[(a)]
Close to the magnetopause, for two configurations 1 and 2, particles have the same behavior if their rigidity  $R_1$ and $R_2$ obey the following relation \citep[][Eqs.~(14)-(16)]{Vogt07}:
\begin{equation}
        \frac{R_1}{R_2}=\left(\frac{\MM_1}{\MM_2}\right)^{1/3}
                        \left(\frac{p_\text{sw,1}}{p_\text{sw,2}}\right)^{1/3}.
        \label{eq-Rscale1}
\end{equation}
\item[(b)]
In contrast, in the inner magnetosphere, for two configurations 1 and 2, two particles have the same trajectory if their rigidity $R_1$ and $R_2$ compare as \citep[][Eq.~(19)]{Vogt07}:
\begin{equation}
        \frac{R_1}{R_2}=\frac{\MM_1}{\MM2}.
        \label{eq-Rscale2}
\end{equation}
\end{itemize}

This different behavior in the two regions can serve as a simple test to see whether the particle shielding is 
dominated by the inner magnetosphere or by the magnetopause.
For this purpose, it is instructive to reproduce Fig. \ref{fig-limits-energy} in terms of the particle ridigity $R$ instead of its energy $E$. Figure \ref{fig-limits-rigidity} thus shows the  magnetospheric filter function $\eta(R,\MM)$ as a function of the particle rigidity $R$ and the planetary magnetic moment $\MM$. Similarly to before, the solid lines represent $R_{\text{10\%}}$ and $R_{\text{90\%}}$, that is,~the particle rigidity for which $\eta=0.10$ and $\eta=0.90$, respectively. 
As a guide to the eye, the dashed lines obey the relation $R\propto \MM$.
The solid lines should follow the dashed lines when Eq. (\ref{eq-Rscale2}) holds, that is,~when the particle shielding is determined in the inner magnetosphere.
Figure \ref{fig-limits-rigidity} shows that, over 
the full
parameter space (e.g.,~for $0.05\MM_\oplus\le\MM\le10.0 \MM_\oplus$), the solid lines (lines of constant $\eta$) approximately follow the dashed lines. This indicates that the magnetic shielding is dominated by the inner magnetosphere, and that the magnetopause contributes only weakly.

\begin{figure}[tb] \begin{center}               
\includegraphics[width=0.95\linewidth]{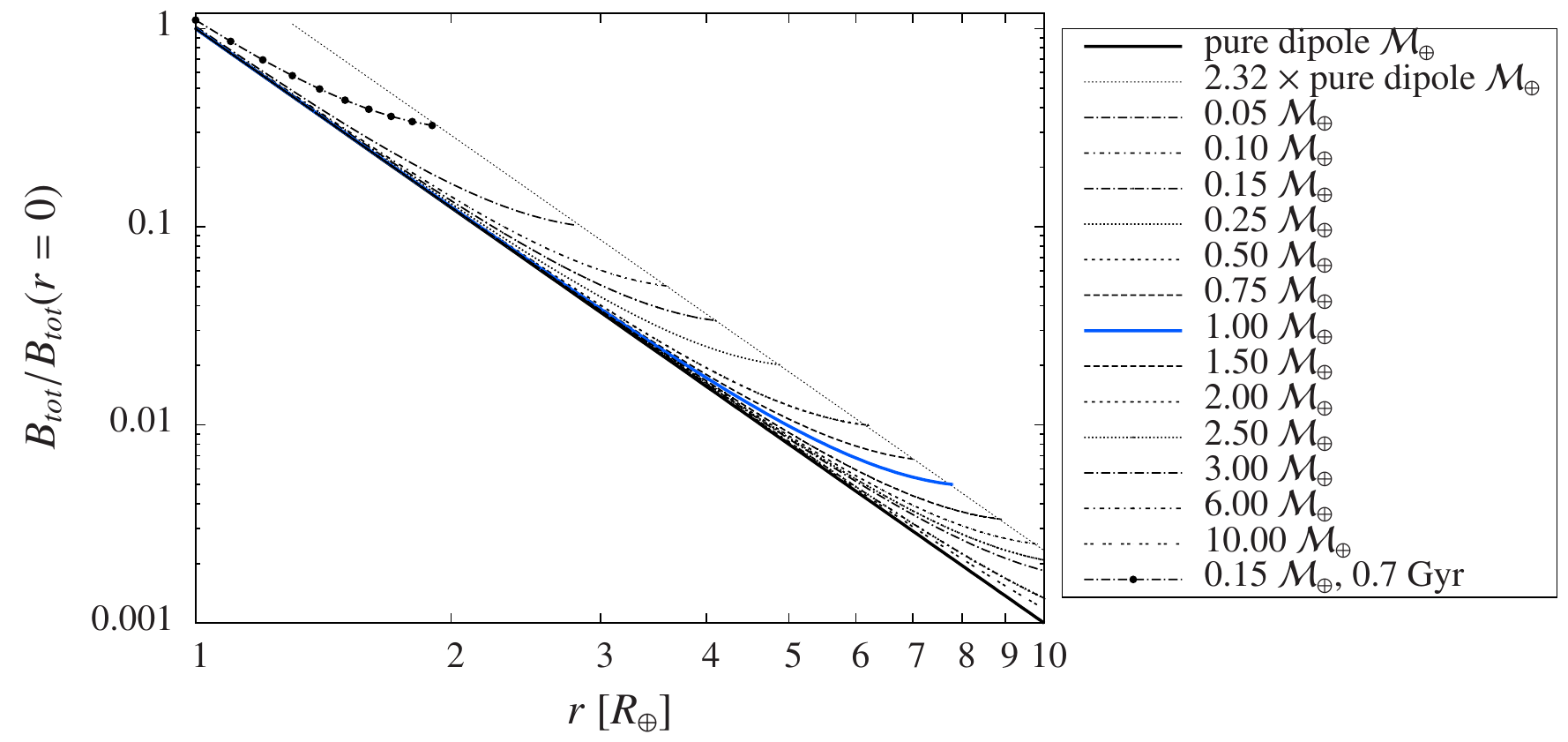}
\caption{Magnetic field between the planetary surface and the substellar point of the magnetopause as a function of distance from the planetary center, relative to the field at the planetary surface.
Thick solid line: pure planetary dipole. Curved blue line: case of a planet with a magnetic moment identical to that of the Earth ($\MM=1.0\,\MM\E$). Curved black lines: magnetic field for different magnetospheric configurations. Straight dotted line: 2.32 times the value for a pure planetary dipole (see text for details).
\label{fig-magfeld-in-simu}}
\end{center}
\end{figure}

This behavior is not surprising: Inspection shows that throughout the magnetosphere, the magnetic field value is close to that of a pure planetary dipole, which decreases as $B\propto r^{-3}$. This is shown in Fig. \ref{fig-magfeld-in-simu}, which shows the magnetic field (relative to the field at the planetary surface) as a function of the distance from the planetary center between the planetary surface and the substellar point of the magnetopause.
The thick solid line represents a pure planetary dipole. The curved lines represent the magnetic field for different magnetospheric configurations (the curved line with dots is discussed below
in more detail). Close to the planetary surface, the magnetic field is identical to the dipole case. At larger distances, the field value differs slightly, but the difference is small.
By construction, the magnetic field value at the magnetopause (defined as the location at which the pressure equilibrium holds, see Sect. \ref{sec-magnetosphere}) is 2.32 times higher than the value for a pure planetary dipole. The locus of these points is shown as a dotted line in Fig. \ref{fig-magfeld-in-simu}. The value of 2.32 results from the magnetospheric form factor of 1.16 (see references in Sect. \ref{sec-magnetosphere}).
However, at that location the magnetic field value is already considerable lower than close to the planetary surface, so that the particle deflection mostly occurs in the inner magnetosphere. This explains why in Fig. \ref{fig-limits-rigidity} the particle shielding is determined in the inner magnetosphere.

It is interesting to compare this to previous results for situations 
where the orbital distance $d$ and the stellar mass $M_{\star}$ were varied \citep[][]{Griessmeier09}.
Four cases were calculated: $d=0.2$ AU for $M_{\star}=0.5 M_{\sun}$, and $d=1.0$ AU for $M_{\star}=1.0 M_{\sun}$, each for two magnetic moment cases ($\MM=0.15\MM_\oplus$ and $\MM=\MM_\oplus$).
Using the same cosmic-ray model as here, it was found that the stellar wind ram pressure (which depends on stellar mass and planetary orbital distance) does
not have a noticeable influence on the cosmic-ray energy spectrum.
In other words, the situation for an exoplanet with $d=0.2$ AU for $M_{\star}=0.5 M_{\sun}$ is virtually identical to that of a planet orbiting the Sun at $d=1.0$ AU.
At that time, this came as a surprise: The stellar wind ram pressure does control the size of the planetary magnetosphere, 
thus could have been expected to influence the magnetic shielding of the planet.
\citet[][]{Griessmeier09} offered two reasons for this behavior:
First, for a constant magnetic moment, a decrease of the size of the magnetosphere is compensated for by an increase of the magnetic field created by the magnetopause currents. Thus, a smaller magnetospheric obstacle has a stronger boundary, while a larger magnetopause generates a weaker contribution to the magnetic field. This compensatory effect would keep the cosmic-ray flux constant when the ram pressure is varied.
Secondly, \citet[][]{Griessmeier09} also concluded that the protons mostly feel the planetary dipole field, and the field of the magnetopause currents has little influence.
In the light of Figs. \ref{fig-limits-rigidity} and \ref{fig-magfeld-in-simu}, we confirm this second argument: 
In the configurations discussed by \citet[][]{Griessmeier09}, the magnetic shielding is dominated by the inner magnetosphere, and the magnetopause region does not contribute much. 
The presence of a magnetopause modifies the magnetic field value close to this boundary layer, but the effect at larger distances is negligible. As most of the magnetic shielding occurs in the inner magnetosphere, the shielding efficiency is not sensitive to the location, or even the existence, of the magnetopause. In other words, the protons mostly feel the planetary intrinsic magnetic dipole field and the field of the magnetopause currents has little influence \citep[cf.][Eq. (20)]{Vogt07}.
This effect alone is sufficient to explain the results for these configurations. The compensatory effect of the magnetic field strength at the magnetopause, while valid, probably has a negligible influence.

\begin{figure*}
\begin{center}
        \subfigure[Cosmic ray impact region (stellar age: 4.6 Gyr).]
        {\includegraphics[angle=0,width=0.45\linewidth]
        {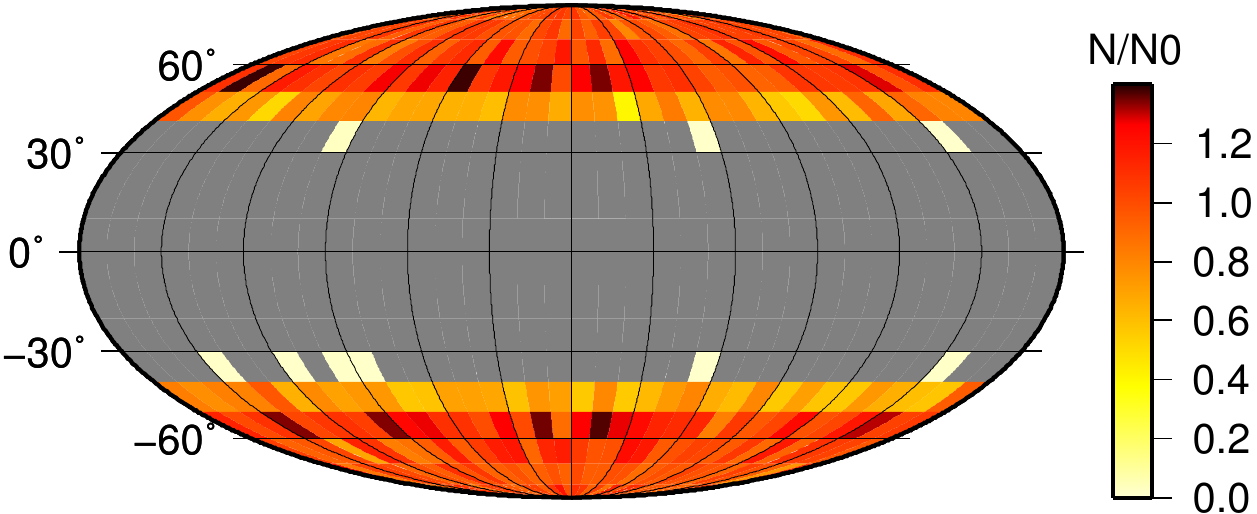}
        \label{fig-anisotropy-07}}
        \subfigure[Cosmic ray impact region (stellar age: 0.7 Gyr).]
        {\includegraphics[angle=0,width=0.45\linewidth]
        {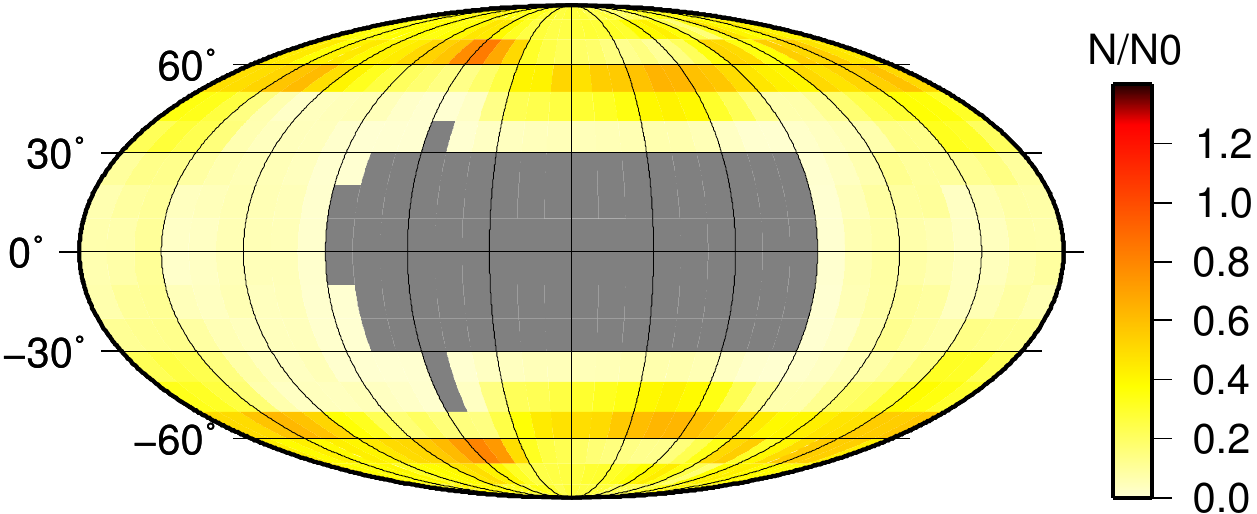}
        \label{fig-anisotropy-46}}
\caption{Cosmic-ray impact region for particles with 256 MeV for the cases presented in \citet[][]{Griessmeier05AB}. Left panel: Weakly magnetized exoplanet ($\mathcal{M}=0.15\, \mathcal{M}_{\oplus}$) at 0.2 AU around an M-dwarf star of 4.6 Gyr age. Right panel:  Weakly magnetized exoplanet ($\mathcal{M}=0.15\, \mathcal{M}_{\oplus}$) at 0.2 AU around an M-dwarf star of 0.7 Gyr age. Gray regions: No particle impact.
\label{fig-anisotropy}}
\end{center}
\end{figure*}

Similarly, we can compare our results to those of \citet[][]{Griessmeier05AB}. In that work, the stellar age $t\sstar$ was varied. This modifies the stellar wind parameters, and thus the size of the planetary magnetosphere. In the case of a very young star ($t\sstar=0.7$ Gyr), the cosmic-ray flux to the planetary surface was indeed modified, and cosmic-ray shielding became slightly more efficient \citep[][Fig. 8]{Griessmeier05AB}. Our Fig. \ref{fig-magfeld-in-simu} (curved line with dots) shows why: Under the action of the very high stellar wind densities and velocities that prevail for young stars, the magnetopause moves to very close distances (less than one planetary radius above the surface). Here, the magnetopause currents do indeed modify the whole magnetosphere by increasing the local magnetic field strength. Even at the surface, the  magnetic field strength is increased by $\sim 10\%$, and throughout a large part of the magnetospheric volume (approximately half), the field strength is increased by at least 40\%. In this case, we have obviously left the regime where particle shielding is determined by the inner magnetosphere. Because of the small magnetosphere, the magnetopause currents have a non-negligible contribution, and cosmic-ray shielding becomes more efficient with increasing stellar wind ram pressure.
This has implications for the anisotropy of low-energy cosmic ray for a planet around a young star.
The magnetic field is more compact on the side of the planet facing the star than on the opposite side. Thus, the flux of Galactic cosmic rays is lower on the star-facing side than on the opposite side. Figure \ref{fig-anisotropy} shows that this is indeed verified in our simulations (0.7 Gyr case), and that this anisotropy does not exist for less compressed situations (4.6 Gyr case). It also shows that the impact area is larger when the shielding is less efficient (0.7 Gyr case vs. 4.6 Gyr case).

\subsection{Cosmic-ray flux energy spectrum at the top of the atmosphere $I(E,\MM)$}
\label{sec-I}

\begin{figure}[tb] \begin{center}
     \includegraphics[width=0.95\linewidth] {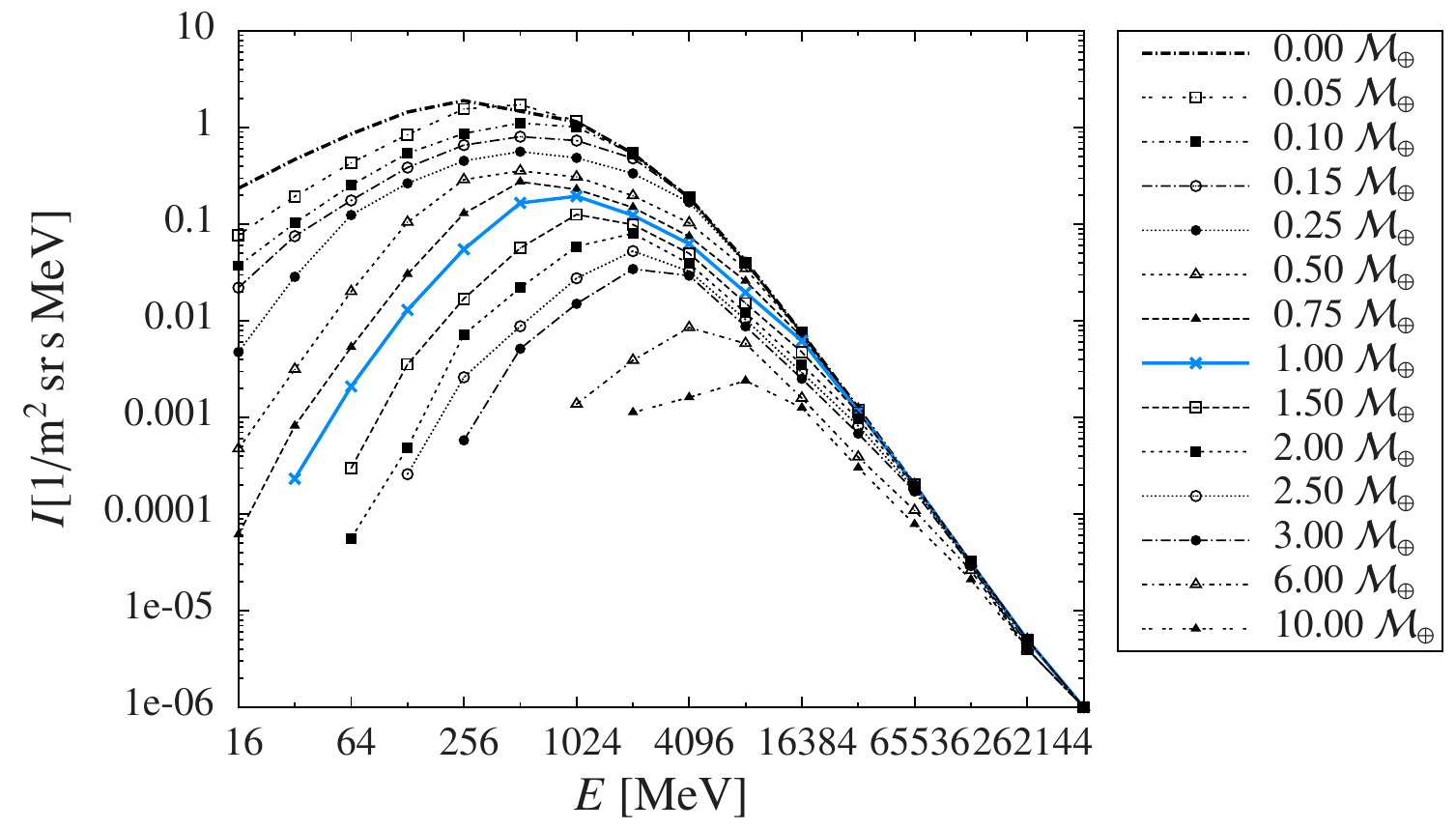}
\caption{Cosmic-ray flux at the top of the atmosphere $I(R,\MM)$ as a function of particle energy and planetary magnetic moment. Dash-dotted line: an unmagnetized planet ($\MM=0$). Blue line: a planet with a magnetic moment identical to that of the Earth ($\MM=1.0\,\MM\E$).
\label{fig-E}}
\end{center}
\end{figure}

To obtain the cosmic-ray energy spectrum $I(E,\MM)$ for extrasolar planets, we first established the cosmic-ray flux outside the magnetosphere. 
For this, we combined the energy spectra from \citet[][]{Seo94} for $E\le$ 8 GeV, and \citet[][using their ``median'' case]{Mori97} for $E\ge$ 16 GeV. This combined dataset represents our reference energy spectrum 
$I_0(E)$, which is shown as a dash-dotted line in Fig. \ref{fig-E}. At the same time, this energy spectrum applies to a non-magnetized planet, that is,~$I_0(E)=I(E,\MM=0)$.

We implicitly assumed a stellar environment similar
to that of the solar neighborhood. 
The cosmic-ray flux could be different, for example, in the case of a different ambient interstellar medium 
\citep{Scherer02,Mueller06,Scherer06,Scherer08}, or in the case of the proximity of the Galactic spiral arms \citep{Scherer06}.
The efficiency of astrospheric shielding (i.e., the fraction of the interstellar cosmic-ray flux penetrating into the astrosphere) also depends on stellar parameters such as the stellar wind \citep{Scherer02,Cohen12}.
 
Strictly speaking, our reference energy spectrum $I_0(E)$ was obtained for an orbital distance of $1.0$ AU. Not all cosmic rays entering an astrosphere reach the inner stellar system.
This is attributed
to effects such as diffusion, convection, adiabatic deceleration, and gradient and curvature drifts \citep[e.g.,][]{Scherer06,Heber06}.
For this reason, the position of the planet within the astrosphere (i.e.,~the orbital distance) can change the particle flux. For distances between 0.1 AU and 1.0 AU 
 \citet[][Sect. 2.5]{Griessmeier09} found that the particle flux varies only weakly ($\sim 30\%$), so that we can neglect this effect for planets in the habitable zones of K/M-dwarf stars.

To obtain the cosmic-ray energy spectrum at the top of the atmosphere for a given magnetosphere configuration, we multiplied the
magnetospheric filter function $\eta(E,\MM)$ (taken from Sect. \ref{sec-eta}, e.g.,~Fig. \ref{fig-eta}) with the cosmic-ray energy 
spectrum outside the magnetosphere $I_0(E)$:
\begin{equation}
        I(E,\MM)=\eta(E,\MM) \cdot I_0(E).
        \label{eq-I}
\end{equation}
Figure~\ref{fig-E} shows the resulting energy spectra at the top of the planetary atmosphere (which we considered to be 100 km above the planetary surface).
The energy spectrum at the top of the atmosphere 
of a planet with a magnetic moment identical to that of the Earth ($\MM=1.0\,\MM\E$) is shown in blue.
Figure~\ref{fig-E}  also contains the previously studied case 
of a weakly magnetized ($\mathcal{M}=0.15\, \mathcal{M}_{\oplus}$) Earth-like exoplanet orbiting 
a K/M-type star with $M_{\star}=0.5 M_{\sun}$ at a distance of $d=0.2$ AU  \citep{Griessmeier05AB,Griessmeier09}. 
Figure \ref{fig-E} shows how this result is generalized for different values of magnetic shielding.
A strong magnetic field reduces the flux of low-energy particles by more than three orders of magnitude, whereas the flux of high-energy particles is barely affected by magnetic shielding.

\begin{figure}[tb] \begin{center}
     \includegraphics[width=0.95\linewidth] {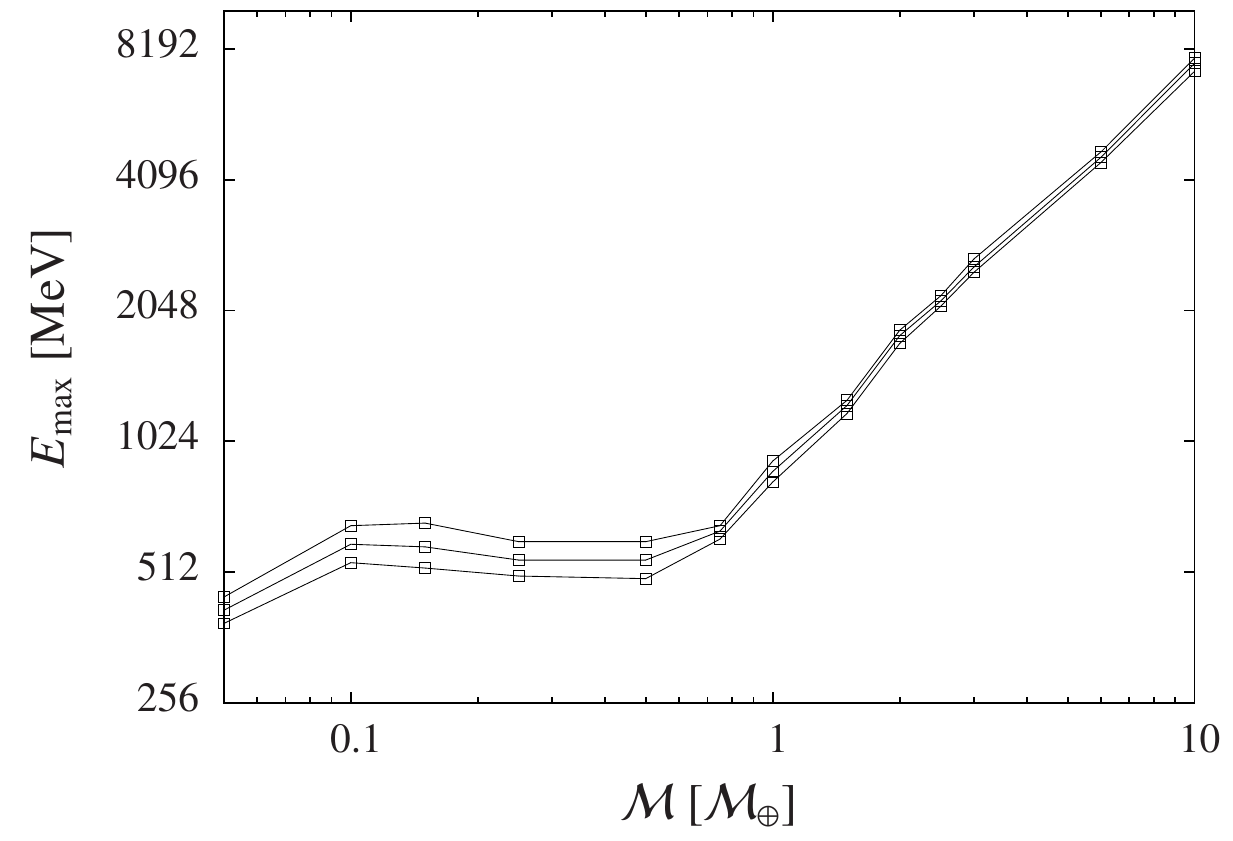}
\caption{Energy with maximum particle flux to the atmosphere as a function of planetary magnetic moment. Middle line: Nominal value. Upper and lower line: Estimated uncertainty.
\label{fig-Emax}}
\end{center}
\end{figure}

The most abundant particle population that reaches the planetary magnetosphere is determined by the peak of the curves in Fig. \ref{fig-E}. The position of this peak (i.e.,~$E_\text{max}=E|_{I=I_\text{max}}$) is plotted against planetary magnetic moment in Fig. \ref{fig-Emax}.
For magnetic moments between 0 and $10\, \mathcal{M}_{\oplus}$, the energy of peak flux increases from $\sim$300 MeV to 8 GeV. 

\begin{figure}[tb] \begin{center}
     \includegraphics[width=0.95\linewidth] {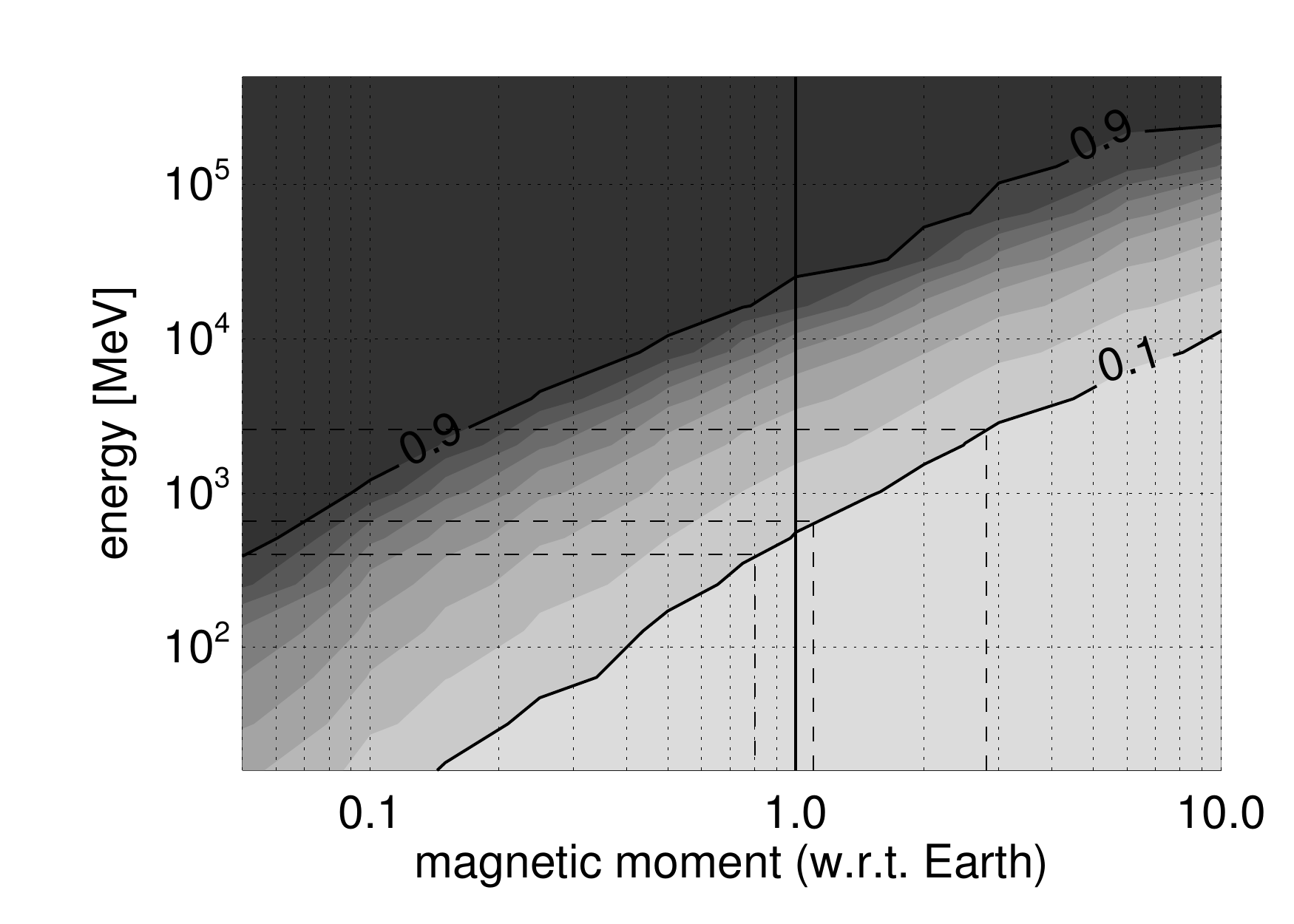}
\caption{As Fig. \ref{fig-limits-energy}, with shielding of cosmic rays at 400 MeV, 660 MeV, and 2.6 GeV. Contours: Magnetospheric filter function $\eta(E, \MM)$ as a function of particle energy $E$ and planetary magnetic moment $\MM$. 
Region below $\eta=0.1$: fully shielded.
Dashed lines: Required magnetic moment to shield cosmic rays of 400 MeV, 660 MeV, or 2.6 GeV.
\label{fig-Elimit}}
\end{center}
\end{figure}

Finally, we return to the comparison between stellar and Galactic cosmic rays. 
Stellar cosmic rays are extremely abundant at low energies, that
is,~below 60 MeV (cf.~Fig. \ref{fig-crspectra}). For a planet with $\MM=0.25\,\MM\E$, these particles are strongly shielded from the atmosphere. Still, even at energies above 60 MeV, stellar cosmic rays can dominate Galactic cosmic rays: 
As discussed in Sect. \ref{sec-populations}, for a planet at 0.2 AU, stellar cosmic rays dominate Galactic cosmic rays for particle energies below 400 MeV, 660 MeV, or 2.6 GeV, depending on stellar activity.
As shown in Fig. \ref{fig-Elimit}, the magnetic moment required to shield particles of these energies are $0.8 \mathcal{M}_{\oplus}$, $1.1 \mathcal{M}_{\oplus}$ and $2.8 \mathcal{M}_{\oplus}$ (we
note that these limits are different for planets at different orbital locations). For strongly magnetized planets 
($\mathcal{M}> 2.8 \mathcal{M}_{\oplus}$), stellar cosmic rays can be entirely neglected, and only Galactic cosmic rays matter. For planets with an intermediate magnetic moment ($0.8 \mathcal{M}_{\oplus} < \mathcal{M} < 2.8 \mathcal{M}_{\oplus}$), stellar cosmic rays are relevant only occasionally, for instance,~during stellar flares, or during periods of high activity. For planets with smaller magnetic moments ($< 0.8 \mathcal{M}_{\oplus}$), both the contribution of stellar and Galactic cosmic rays matter. As stellar cosmic rays can have a non-negligible contribution for planets within
 a wide range of planetary magnetic moments, their influence on exoplanetary atmospheres has indeed to be studied; this is done in a companion article \citep{TabatabaVakili15}.

\section{Conclusion}\label{sec-conclusions}

Magnetic fields on most super-Earths around M-dwarf stars are likely to be weak and  short-lived in the best case, or even non-existent in the worst case.
With this in mind, the question of planetary magnetic shielding against Galactic cosmic rays becomes important.
Instead of trying to estimate the planetary magnetic moment quantitatively, we calculated a large number of representative cases and systematically studied the influence of the planetary magnetic field on the flux of Galactic cosmic rays to the planet.
At the highest energies, we found that the flux of Galactic cosmic-ray particles is barely affected by magnetic shielding. For lower particle energies, however, we found that the particle flux to the planetary atmosphere can be increased by more than three orders of magnitude in the absence of a protecting magnetic field. 
Between unmagnetized and strongly magnetized planets, the maximum energy for partial shielding increases from 512 MeV to 200 GeV. 
In the atmosphere, unshielded energetic particles can destroy 
atmospheric ozone and other biomarker molecules.
Implications include a modification of the planetary emission and transmission spectrum and an enhanced surface UV flux. In
addition, secondary muons can reach the planetary surface, leading to a high biological dose rate.
These effects will be discussed in paper II. 
For a planet with $\MM \le 0.25\,\MM\E$, low-energy stellar cosmic rays can have a strong effect that may even dominate Galactic cosmic rays.
This case
is analyzed in a companion article \citep{TabatabaVakili15}.

\begin{acknowledgements}

This study was supported by the International Space Science
Institute (ISSI) and benefited from the ISSI Team 
``Evolution of Exoplanet Atmospheres and their Characterisation''.
The cosmic-ray simulations for this work were performed on the Clairvaux cluster, using
1984 days of CPU time. Special thanks to I. Cognard and G. Desvignes.
We would also like to thank the referee for his or her excellent comments and suggestions that
helped to improve the paper.

\end{acknowledgements}


\bibliographystyle{elsart-harv}

\end{document}